\documentclass[a4paper,fleqn,usenatbib]{mnras}

\usepackage{newtxtext,newtxmath}
\usepackage[T1]{fontenc}
\usepackage{ae,aecompl}

\usepackage{graphicx}	% Including figure files
\usepackage{amsmath}	% Advanced maths commands
\usepackage{amssymb}	% Extra maths symbols

\title[Dark energy EOS from lensed GWs]{Complementary constraints on dark energy equation of state from strongly lensed gravitational wave}

\author[Liu et al.]{
Bin Liu,
Zhengxiang Li,\thanks{E-mail: zxli918@bnu.edu.cn}
and Zong-Hong Zhu
\\
Department of Astronomy, Beijing Normal University, Beijing 100875, China\\
}

\begin{document}
\label{firstpage}
\pagerange{\pageref{firstpage}--\pageref{lastpage}}
\maketitle

\begin{abstract}
It has been shown that time delay of strong gravitational lensing is not only an effective cosmological probe to constrain the Hubble constant, the matter density and the curvature of the universe but also useful for breaking the degeneracy of the dark energy equation of state and thus provide complementarity with other popular probes, such as type Ia supernovae, baryon acoustic oscillations, and cosmic microwave background radiation. Interestingly, compared to traditional strong lensing systems where quasars act as sources, strongly lensed gravitational waves (GWs) from compact binary coalescence and their electromagnetic (EM) counterparts systems have been recently proposed to be more powerful for studying cosmology since GWs and their EM counterparts are transients. Short durations of GWs and their EM counterparts are very advantageous to time delay measurement and lens profile modeling. Here, in the framework of Chevalier-Polarski-Linder (CPL) parametrization, we investigate improvement of constraining power on dark energy equation of state due to including time delay measurements of strong lensed GW systems. It is suggested that, on the basis of the third generation ground-based detector, e.g. Einstein Telescope, adding time delay of only 30 strong lensed GW systems to type Ia supernovae and cosmic microwave background radiation can improve the dark energy figure of merit by a factor 2. For the traditional standard siren method where the uncertainties of luminosity distances of GWs are $\sim10\%$, a few $\times~10^4$ events are expected to present similar constraints. In the precision cosmology era, this progress is of great significance for studying the nature of dark energy.
\end{abstract}


%%%%%%%%%%%%%%%%%%%%%%%%%%%%%%%%%%%%%%%%%%%%%%%%%%

%%%%%%%%%%%%%%%%% BODY OF PAPER %%%%%%%%%%%%%%%%%%

\section{Introduction}

Time delay of strong lensing system is considered to be a powerful cosmological probe. It can be used as an independent method to get the information of cosmological parameters. It was demonstrated that we can use time delay to determine the Hubble constant in an independent way \citep{resfdal1964}. Recently, \cite{suyu2010} showed that utilizing the time delay of lensing events to constrain the Hubble constant with precision competitive to other methods is possible. With the increase number of the lensing events, for example, the upcoming Large Synoptic Survey Telescope(LSST) will detect 3000 well-measured lensing time delays in 8000 lensed quasars\citep{oguri2010}, the results will be further improved. And \citet {liao2017} also presented that time delay can constrain cosmic curvature very well with the upcoming observations of LSST.
		
		It has been demonstrated that the universe is undergoing an accelerated expansion \citep{riess1998, perlmutter1999}, so an exotic component dubbed dark energy with negative pressure was introduced to account for this unexpected discovery. Phenomenologically, dark energy is a homogeneous fluid described by the equation of state $w=p /\rho$. Any accuracy result of dark energy equation of state is of great importance to understand the physical mechanism behind. Time delay of strong lensing system has also been proposed as an effective probe to study the equation of state of dark energy.  \citet{linder2004, linder2011} pointed out that lensing time delay can provide complementarity with other prevalent cosmological probes to constrain cosmological parameters. The combination of three angular diameter distance named time delay distance derived from time delay measurements and lens profile modeling has different degeneracy directions from those of other popular probes such as type Ia supernovae (SNe Ia), cosmic microwave background (CMB) radiation. More specifically, for the most popular Chevalier-Polarski-Linder (CPL) parametrization \citep{chevalier2001, linder2003}, $w_0$ and $w_a$ will have positive correlation instead of anticorrelation in other probes when lenses lie in the redshift range $0-0.6$. It was suggested that the dark energy figure of merit can be improved by a factor of 5 by adding 150 "golden lens" where precise reconstructions of the lens mass model is assumed to be achieved in the future observation to the combination of SNe Ia and CMB observations available at that time \citep{linder2011}. The result is comparable with the latest constraints from the combination of CMB, SNe Ia, and baryon acoustic oscillations (BAO) information~\citep{planck2018}. 
		
		With the detection of gravitational waves (GW) from the merger of double compact object (DCO) \citep{abbott2016a, abbott2016b, abbott2017a, abbott2017b, abbott2017c}, we come into a new era of astronomy. Gravitational waves provide us with a completely new means of observation and are also a promising probe for cosmology. For instance, \citet{schutz1986} proposed that the luminosity distance obtained independently from the gravitational wave signal can be used to constrain the Hubble constant by combining the redshift information of source. Therefore, gravitational wave signals from the merger of DCOs are put forward as distance indicators and are called standard sirens. Results from these distance measurement can be used to crosscheck with other probes.
		
		Lensing effect of GW signals have been widely explored in the literature \citep{wang1996,nakamura1998,takahashi2003,cao2014} and Laser Interferometer Space Antenna (LISA) also prepares to use the time delay of lensed GW for studying cosmology \citep{sereno2011}. They showed that the Hubble constant can be constrained to $\sim 10\%$. The detection of the GW signals from binary neutron stars \citep{abbott2017c} implied that we can not only detect the GW signals, the observations of its EM counterparts is also possible. Recently, \citet{liao2017} proposed future strongly lensed GW and the corresponding EM signals as a precision cosmological probe. Unlike the traditional lensed quasar systems, the transient nature of the merger events of GW source, i.e. DCOs, ensures that the time delay can be measured to great precision. If we can locate the host galaxy by means of EM counterparts, redshift information of the GW source can be easily obtained. Moreover, the precision and accuracy of lens modeling for strong lensed GW systems can be significantly improved since there is no dazzling AGN contamination in the host galaxy. It was obtained that 10 such events will constrain the Hubble constant to $0.68\%$ in the $\Lambda$CDM model \citep{liao2017}. Meanwhile, this promising system has been proposed for extensive studies, such as testing the speed of GW \citep{fan2017, collett2017}, inducing time delays in multi-messenger signals from the same source to probe fundamental physics \citep{baker2017}, acting as a powerful cosmic ruler \citep{wei2017}, testing the weak equivalence principle \citep{yu2018}, and probing dark matter substructures \citep{liao2018} and gravity \citep{yang2018}.
		
		Although only a few GW events have been detected, the next generation detectors will improve their detection ability significantly. For example, the third-generation ground-based detector, i.e. Einstein telescope (ET), will expand the detection space by three orders of magnitude, and thus can detect much more GW events. It was shown that about $10^4$-$10^5$ GW events will be detected by ET per year and 50-100 of them are likely to be strongly lensed \citep{piorkowska2013,biesiada2014,ding2015}. This expected considerable number of lensed GW events implies that it is possible to use these systems for estimating cosmological parameters. Here, we study the ability of the lensed GW+EM signals for constraining the dark energy equation of state.
		
		In this paper, we take the flat $\Lambda$CDM universe as our fiducial model in the simulation. The matter density parameter $\Omega_m=0.315$ and the Hubble constant $H_0=67.3~\mathrm{km/s/Mpc}$ from the latest $Planck$ CMB observations~\citep{planck2018} is taken for Monte Carlo simulations in our analysis.

\section{Methodology}

       Strong gravitational lensing is a robust tool in astrophysics and cosmology \citep{treu2010}. A strong gravitational lensing system has multiple images and these signals take different time to reach the observer. Time delay consists of two contributions, one is caused by different paths of light travelling through, the other is the difference of lens potentials at different positions through the gravitational field of the lens. According to the gravitational lensing theory, the total time delay is
		\begin{equation}
		\Delta t_{i,j}=\frac{(1+z_{l})D_{\Delta t}}{c}\Delta \phi_{i,j},
		\end{equation}
		where $\Delta t_{i,j}$ is the time delay between two images or GW signals, $z_l$ is the redshift of lens and $c$ is the light speed or the GW speed. The time delay distance, $D_{\Delta t}$, is a multiplicative combination of the three angular diameter distances,
		\begin{equation}
		D_{\Delta t}=\frac{D_l(z_l)D_s(z_s)}{D_{ls}(z_l,z_s)},
		\end{equation}
		where $z_s$ is the redshift of the source. $D_l$, $D_s$, and $D_{ls}$ are angular diameter distances from the lens to the observer, from the source to the observer, and from the lens to the source, respectively. $\Delta \phi$ is the difference of Fermat potential of $i$, $j$ images and it can be written as
		\begin{equation}
		\Delta \phi_{i,j}=\frac{(\theta_{i}-\beta)^2}{2}-\Psi(\theta_{i})-\frac{(\theta_{j}-\beta)^2}{2}+\Psi(\theta_{j}),
		\end{equation}
		where $\theta_{i,j}$ is the angular position of $i$, $j$ images and $\beta$ is the angular position of the source. $\Psi$ is the two-dimensional lens potential. For the traditional lensed quasars, one has to monitor light curves for a long time to measure time delay between images. For lensed GWs, it is only necessary to look for signals which have the same duration, frequency drift, amplitude change rate and come from the similar sky position. The only difference is the amplitude of the signals because of the magnification effect of the lens.
		
		In the flat FLRW framework, the angular diameter distance can be expressed as,
		\begin{equation}
		D(z;\mathbf{p})=\frac{1}{1+z}\frac{c}{H_0}\int_0^z \frac{dz'}{E(z';\mathbf{p})},
		\end{equation}
		where $H_0$ is the Hubble constant, $E(z';\mathbf{p})=H(z)/H_0$ is the dimensionless expansion rate, $\mathbf{p}$ is a set of cosmological parameters.
		When we take the CPL parametrization proposed by \citet{chevalier2001} and \citet{linder2003} to characterize the dark energy evolution with respect to redshift, where
		\begin{equation}
		w(z)=w_0+w_a \frac{z}{1+z},
		\end{equation}
		the expansion rate can be written as
		\begin{equation}
		E^2(z';\mathbf{p})=\Omega_m(1+z)^3+(1-\Omega_m)(1+z)^{3(1+w_0+w_a)}\exp\bigg(\frac{-3w_az}{1+z}\bigg).
		\end{equation}
		In this case, there is a positive correlation between $w_0$ and $w_a$ in the time delay distance in the redshift range $0<z_l<0.6$ \citep{linder2004}, which is orthogonal to some other popular probes where $w_0$ and $w_a$ are negatively correlated. This feature ensures that time delay distance can provide complementary constraints on the dark energy equation of state to some other currently popular probes.
		
		From the observation of lensing events, we can measure the time delay between the lensed signals and we can identify the host galaxy of the source and obtain high-resolution images. This will help us to get the precise and accuracy information about the Fermat potential. Then, the time delay distance is derived and can be applied to infer cosmological information via the likelihood function $L \sim e^{-\chi^2/2}$, 
		\begin{equation}
		\chi^2= \sum_i\frac{\bigg[D_{\Delta t}^{\rm th}(z_l, z_s; \mathbf{p})-D_{\Delta t,i}^{\rm obs}\bigg]^2}{\sigma^2_{D_{\Delta t,i}^{\rm obs}}},
		\end{equation}
		where $\sigma_{D^{\rm obs}_{\Delta t}}$ is the observational uncertainty of the time delay distance mainly contributing from three ingredients, the uncertainty from measurements of time difference between images $\sigma_{\Delta t}$, the uncertainty from measurements from Fermat potential difference between paths $\sigma_{\Delta \Phi}$, and the systematic error from the mass distribution along the line of sight $\sigma_{\rm LOS}$. The total time delay distance error of each strong lensing system is 
       	\begin{equation}
       	\sigma^2_{D^{\rm obs}_{\Delta t}}=\sigma^2_{\Delta t}+\sigma^2_{\Delta \Phi}+\sigma^2_{\rm LOS}.
       	\end{equation}
		 
		 At present, observations of strong lensing time delay are in EM bands (mainly in optical band) and only a small number of lensing systems are well measured. By monitoring light curves of quasars for traditional strong lensing systems, the average relative uncertainty from time delay measurements, $\delta_{\Delta t}=\sigma_{\Delta t}/\Delta t$, is about $3\%$ shown by the first time delay challenge (TDC1) \citep{liao2015}. The average relative uncertainty of Fermat potential difference or lens modeling, $\delta_{\Delta \Phi}=\sigma_{\Delta \Phi}/\Delta \Phi$, is also about $3\%$ due to contaminations of dazzling AGNs in the center of the source. For the intractable systematic error caused by the contaminations of mass distributed along the line of sight, the state-of-the-art studies suggest that this component, $\delta_{\rm LOS}$, can reach the level of $2\%$ \citep{bonvin17,cristian18,tihhonova18}. Fortunately, future detections of lensed GWs will significantly improve the present situation. First, the time delay in strong lensed GW systems can be accurately determined ($\delta_{\Delta t}\sim0$) owing to the small ratio between the short duration of GW signals from the merger of DCOs, $\sim \mathcal{O}(10^{-1}~{\rm s})$, and the typical galaxy-lensing delay time $\sim\mathcal{O}$(10 days). Second, high quality images of the host galaxy can be obtained since there is no central dazzling AGN. This advantage will improve the precision of lens modeling by a factor of $\sim4$ \citep{li2018}. That is, the relative error of Fermat potential reconstruction ($\delta_{\Delta \Phi}$) can be reduced to $\sim 0.8\%$. For the systematic error from the contaminations of mass distributed along the line of sight, lensed GW systems face the same difficulties as lensed quasar.

\section{simulation and results}
        In addition to the above-mentioned uncertainty levels for three ingredients, we also should clarify the redshift distributions of lens and source in strongly lensed GW systems to estimate the constraining power of future lensed GW on dark energy evolution. First, we calculate the redshift distribution of the source of lensed GW systems detected by ET. The cumulative yearly detection of the lensed GW events up to the redshift $z_s$ can be written as \citep{biesiada2014},
		\begin{equation}
		\dot{N}_{\mathrm {lensed}}(z_s)=\int_0^{z_s}\tau(z_s)\frac{d\dot{N}(>\rho_0)}{dz}dz.
		\end{equation}
		$\tau$ is the total optical depth:
		\begin{equation}
		\tau(z_s) = \frac{16}{30}\pi^3\frac{c}{H_0}\widetilde{r_s}^3\bigg(\frac{\sigma_*}{c}\bigg)^4n_*\frac{\Gamma\bigg(\frac{4+\alpha}{\beta}\bigg)}{\Gamma\bigg(\frac{\alpha}{\beta}\bigg)}y_{\mathrm{max}}^2,
		\end{equation}
		where $\widetilde{r}=\int_0^z\frac{dz'}{E(z')}$ is the dimensionless comoving distance, and $\sigma_*=161\pm5~\mathrm{km/s}$, $n_*=8.0\times10^{-3}h^3~\mathrm{Mpc}^{-3}$, $\alpha=2.32\pm0.10$, $\beta=2.67\pm0.07$ are the parameters of velocity dispersion distribution function of elliptical galaxies and their value is taken from \cite{choi2007}. $\frac{d\dot{N}(>\rho_0)}{dz}dz$ is the intrinsic detection rate with the signal-to-noise (SNR) exceeding the detector's threshold (we take the threshold $\rho_0=8$ in our work):
		\begin{equation}
		\frac{d\dot{N}(>\rho_0)}{dz}=4\pi\bigg(\frac{c}{H_0}\bigg)^3\frac{\dot{n_0}(z_s)}{1+z_s}\frac{\widetilde{r_s}^2(z_s)}{E(z_s)}C_{\Theta}(x(z_s,\rho_0)).
		\end{equation}
		$\dot{n}(z_s)$ is the DCO inspiral rate of each redshift calculated by \citet{dominik2013}. $C_{\Theta}(x(z_s,\rho_0))$ is the function determined by the detector's sensitivity and orientation. In their calculation, two galaxy metallicity evolution with redshift were assumed, that is, the "high end" and "low end" situation. Moreover, four DCO formation scenarios: standard scenario, optimistic common envelope scenarios, delayed SN explosion scenarios and high BH kick scenario were considered.
		
\begin{figure}
			\centering
			\includegraphics[scale=0.5]{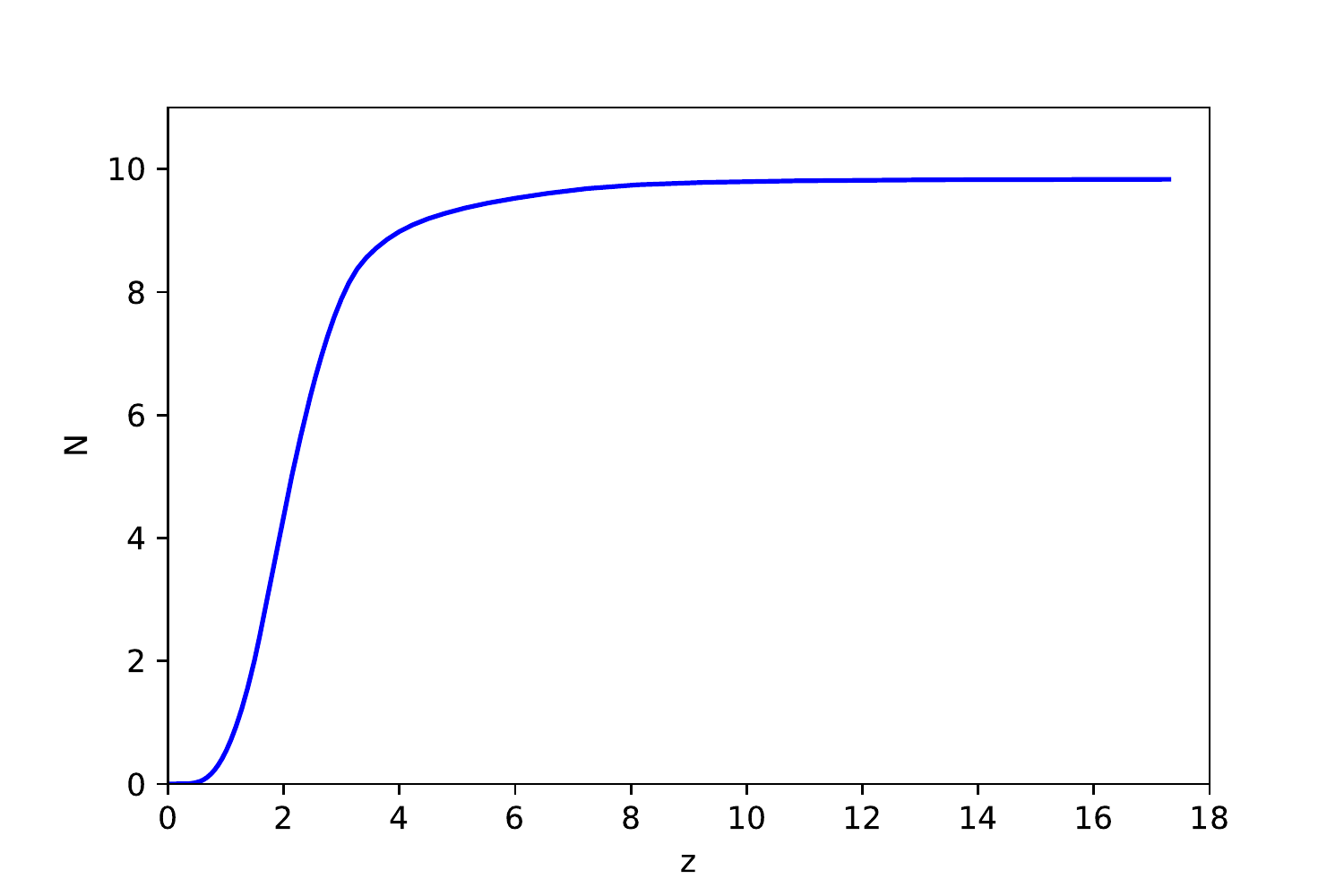}
			\caption{The cumulative redshift distribution of strongly lensed GW events with sources being NS-NS and NS-BH mergers yearly detected by ET within redshift $z$. Both events with the weaker image being $\rho_{\mathrm{int}}<8$ and $\rho_{\mathrm{int}}>8$ are included.}
			\label{events}
		\end{figure}

		Owing to the magnification effect of the lens, some GW events with weak intrinsic signals, that is, their SNR $\rho_{\rm int}$ is smaller than the detector threshold $\rho_0$, can be fortunately detected after lensing magnification, the detection rate of these events is,
		\begin{equation}
		\frac{\partial^2\dot{N}}{\partial z_s\partial \rho}=4\pi\bigg(\frac{c}{H_0}\bigg)^3\frac{\dot{n}(z_s)}{1+z_s}\frac{\widetilde{r}^2(z_s)}{E(z_s)}P_{\Theta}(x(z_s,\rho))\frac{x(z_s,\rho)}{\rho},
		\end{equation}
		and then the detection rate of these lensed GW systems is :
		\begin{equation}
		\dot{N}_{\mathrm{lensed}}(z_s)=\int_0^{z_s}dz_s\int_0^{\rho_0}\tau(z_s,\rho)\frac{\partial^2\dot{N}}{\partial z_s\partial \rho}d\rho,
		\end{equation}
		(see \cite{ding2015} for more details). The total detection rate includes both the case with the weaker image $\rho_{\mathrm{int}}<8$ (but $>8$ after magnification) and the one with the stronger image $\rho_{\mathrm{int}}>8$. By adding up these two detection rates, the probability distribution function (PDF) of the redshift of the source can be obtained. Here, electromagnetic counterparts are assumed to have their redshifts measured. Therefore, NS-NS and NS-BH events are considered in our analysis. Figure \ref{events} shows their cumulative redshift distribution of these two kinds of source.  On the whole, $\sim \mathcal{O}(10)$ strongly lensed GW systems together with their EM counterparts might be registered by ET and follow-up facilities. It is reasonable to expect that tens of these interesting systems could be collected by third generation ground-based detector within 5-10 years. We mock GW source redshifts $z_s$ which satisfy this distribution for our following analysis. 
				
\begin{figure}
    \centering
    \includegraphics[scale=0.2]{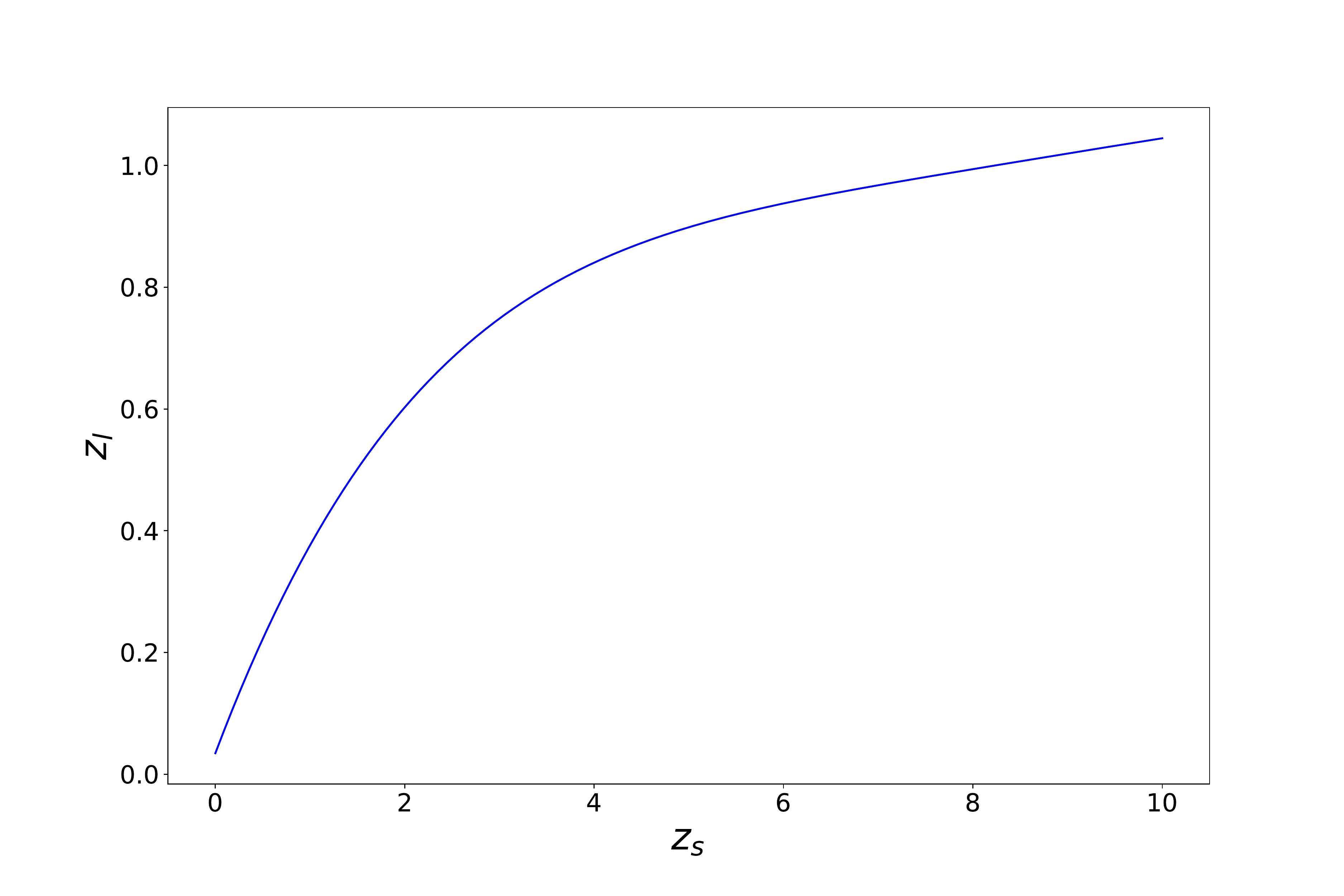}
    \caption{The lens redshift $z_l$ to produce the maximal differential lensing probability as a function of the source redshift $z_s$.}
    \label{zszl}
    \end{figure}

     We assume that it is possible to identify the lens galaxy with the help of optical or near infra-red facilities after ET registering a lensed GW event. In our simulation, we draw the most likely redshift of the lens for a lensed GW+EM event with the source at given redshift $z_s$. The differential optical depth for an elliptical galaxy can be expressed as
    \begin{equation}
    \frac{d\tau}{dz_l}=16\pi^3\bigg(\frac{c}{H_0}\bigg)^3\frac{\widetilde{r}_{ls}^2\widetilde{r}_l^2}{\widetilde{r}_s^2E(z_l)}y_{\mathrm{max}}^2n_*\frac{\Gamma\bigg(\frac{4+\alpha}{\beta}\bigg)}{\Gamma\bigg(\frac{\alpha}{\beta}\bigg)}.
    \end{equation}
     For a source at redshift $z_s$, we can calculate the differential optical depth of any $z_l$ at $0<z<z_s$. In this range, there is a $z_l$ that produces the maximal differential lensing probability. Figure \ref{zszl} shows the lens redshift $z_l$ maximizing the differential lensing probability as a function of the source redshift $z_s$. Therefore, for a sample of lensed GW systems with their source redshifts satisfying the distribution shown in Figure \ref{events}, we pick each system one by one and then individually determine its lens redshift by considering $z_s-z_l$ curve presented in Figure \ref{zszl}.

		In order to investigate the complementarity of strongly lensed GW systems to already existing popular probes on constraining dark energy equation of state, we include the latest $Planck$ 2018 CMB observations \citep{planck2018, lu2018} and SNe Ia from the upcoming DES Hybrid 10-field Survey(Table 14 of  \citet{bernstein2012}. For CMB data, we use the information about the heights of the peaks of CMB temperature  spectrum along the line-of-sight, i.e. the shift parameter $R$. For SNe Ia, the mock distance modulus-redshift data given in \citet{bernstein2012} is used. The parameters set is $\{\Omega_m,H_0,w_0,w_a\}$. We use the minimization function to get the parameters sets of minimum $\chi^2$ and run the simulation 10000 times with different random seeds.
		
\begin{figure}
			\centering
			\includegraphics[scale=0.6]{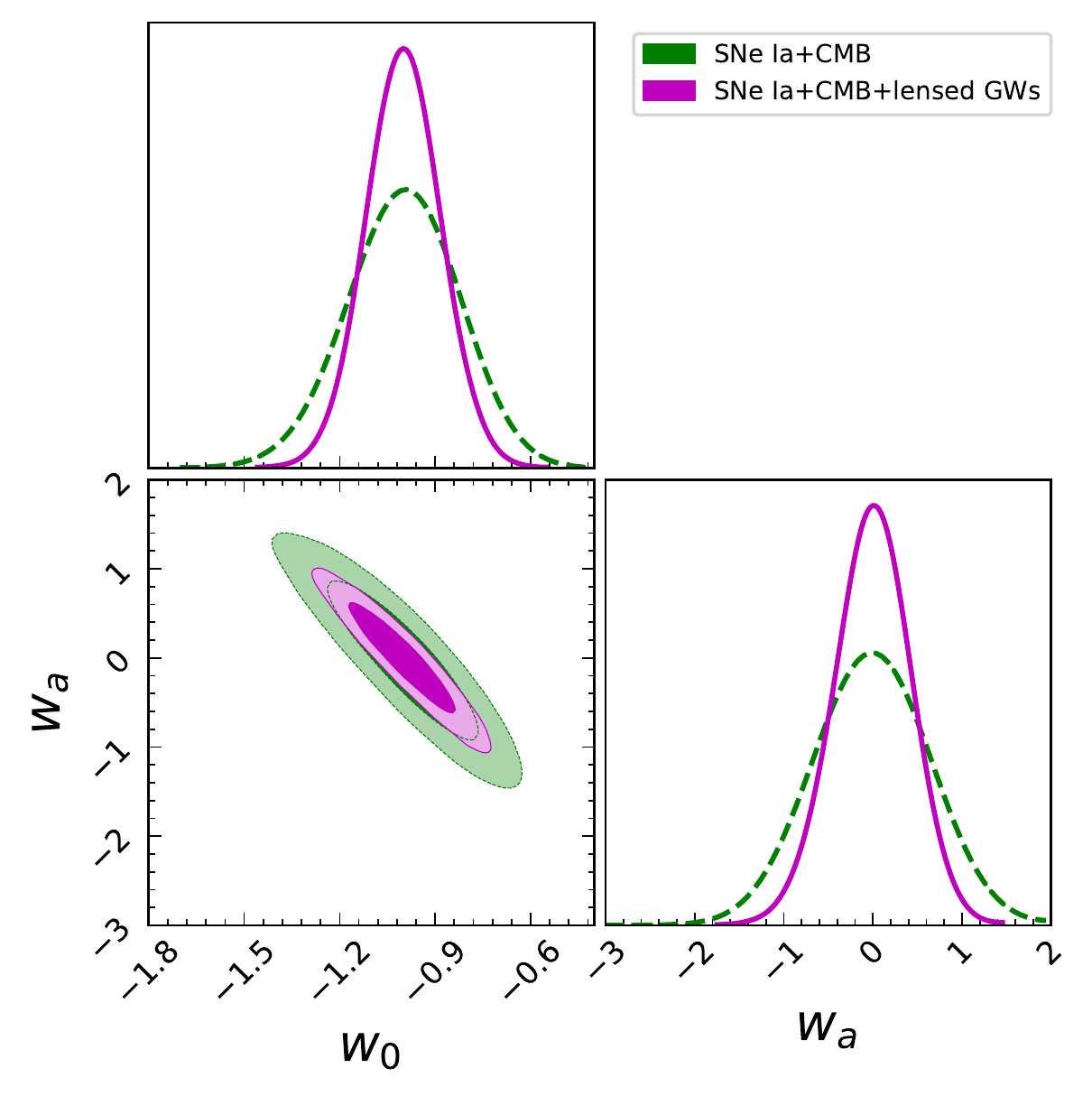}
			\caption{Marginalized PDFs and the $68\%,~95\%$ confidence contours of the dark energy parameters $w_0$ and $w_a$,   SNe Ia and CMB, plus 10 lensed GW systems are considered.}
			\label{10GW and EM signals}
		\end{figure}

\begin{figure}
				\centering
				\includegraphics[scale=0.6]{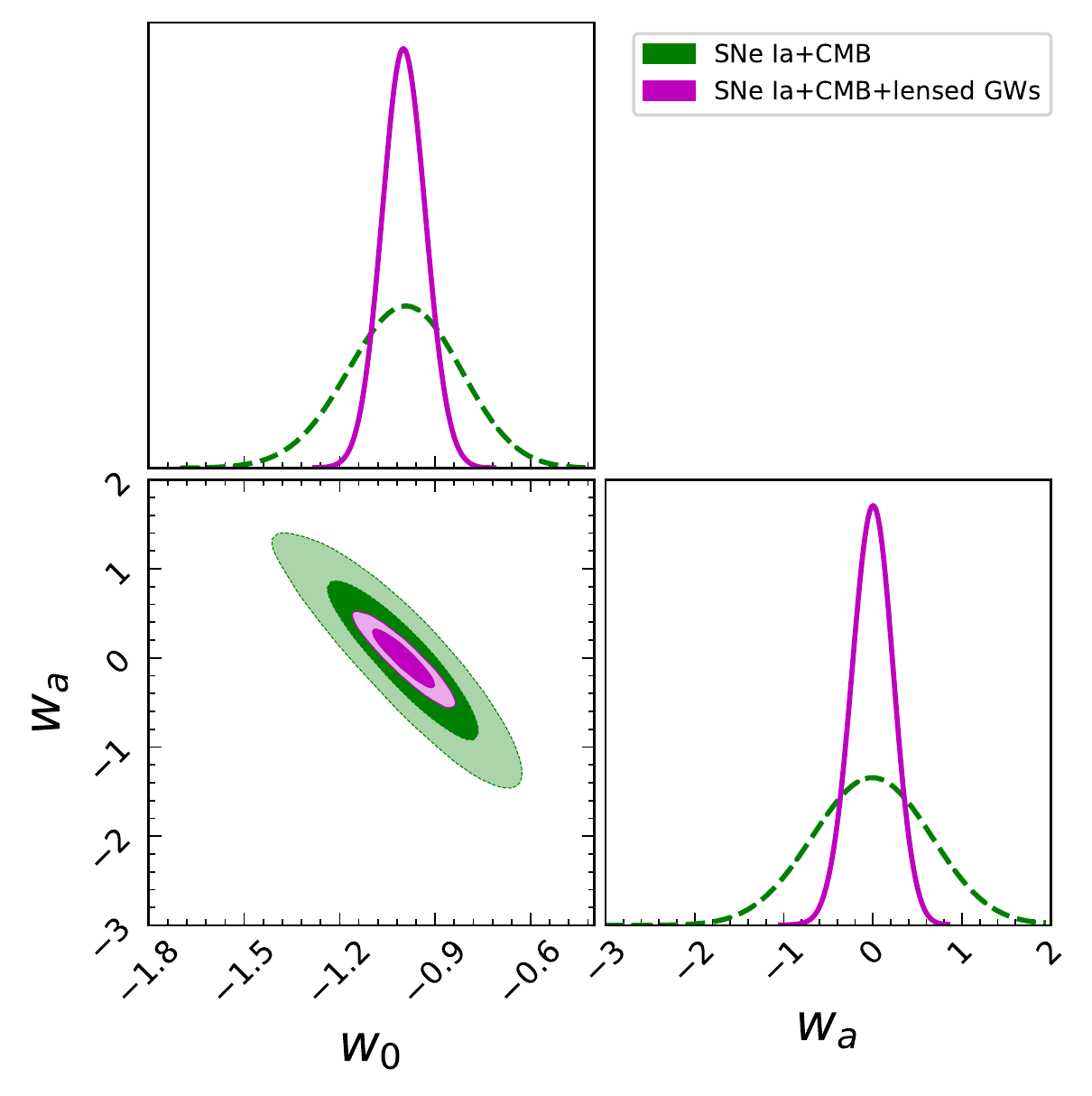}
				\caption{Marginalized PDFs and the $68\%,~95\%$ confidence contours of the dark energy parameters $w_0$ and $w_a$, SNe Ia and CMB, plus 100 systems are considered.}
				\label{100GW and EM signals}
			\end{figure}

\begin{figure}
				\centering
				\includegraphics[scale=0.3]{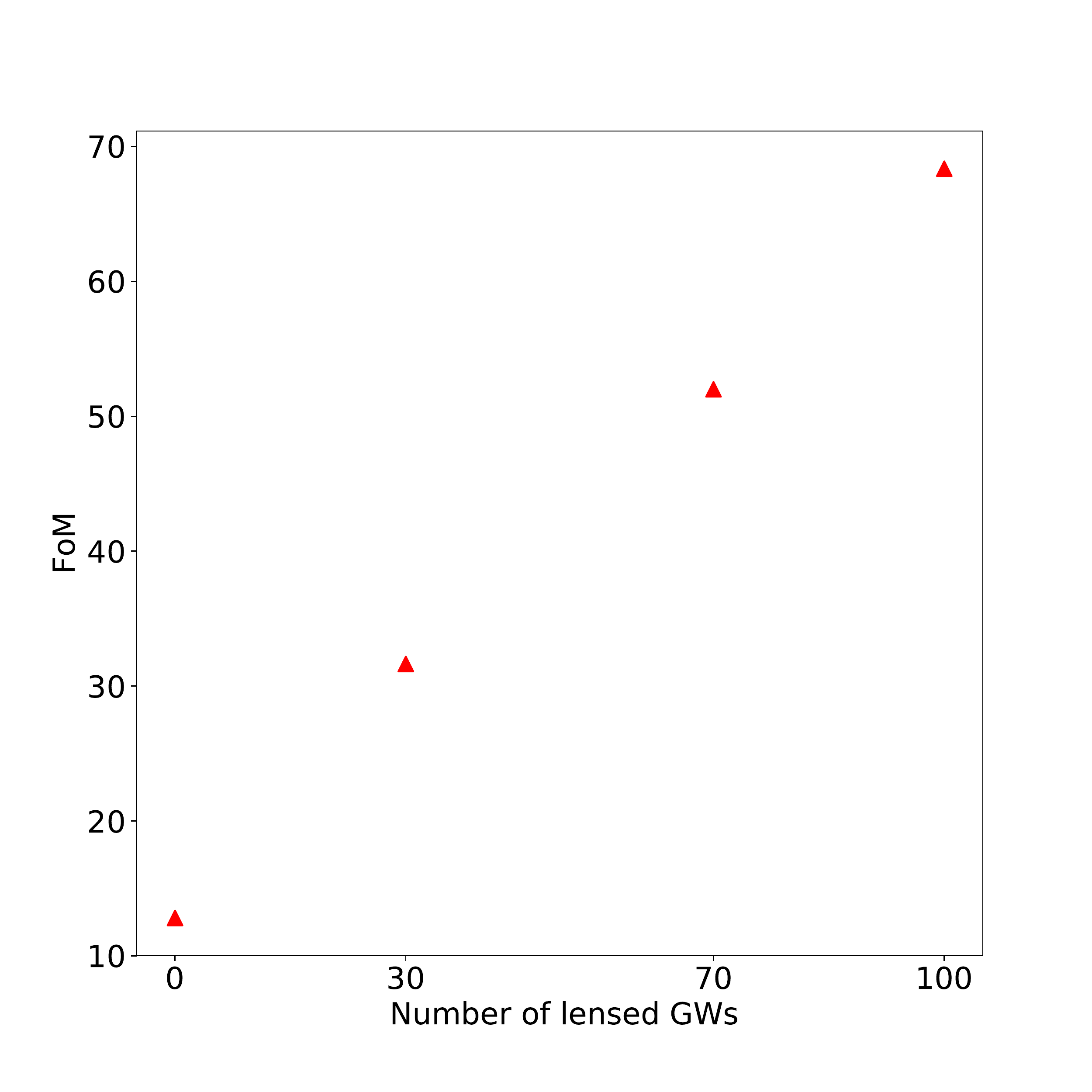}
				\caption{The figure of merit (FoM) of constraint on the dark energy equation of state from currently available popular probes (SNe Ia+CMB) plus different number of time delay measurement of strongly lensed GW systems.}
				\label{FoM}
			\end{figure}

		We consider several tens of lensed GW and EM events which are likely to be collected by ET to estimate the complementary constraints on $w_0$ and $w_a$. Results are shown in Figures (\ref{10GW and EM signals}, \ref{100GW and EM signals}) and Table \ref{tab1}. In order to quantify the improvement of constraints on dark energy equation of state after taking lensed GW systems into account, we apply the figure of merit (FoM) \citep{albrecht2006,wang2008,dossett2011,sendra2011} which is proportional to the inverse area of the error ellipse in the $w_0-w_a$ plane, 
		\begin{equation}
		\mathrm{FoM}=[\mathrm{det}C(w_0, w_a)]^{-1/2}, 
		\end{equation}
		where $C(w_0, w_a)$ is the covariance matrix of $w_0-w_a$ after marginalizing over all other cosmological parameters. Larger FoM implies stronger constraint on the parameters since it relates to a smaller error ellipse. FoMs of constraint on the dark energy equation of state from currently available $Planck$ 2018 CMB observations and upcoming DES SNe Ia together with different number of time delay measurement of strongly lensed GW systems are plotted in Figure \ref{FoM}. It is suggested that the constraining power is improved by a factor of 2 if only 30 strongly lensed GW systems are considered jointly with other popular probes. In the era of precision cosmology, constraints from observations of almost all popular probes are consistently in the favor of the standard $\Lambda$CDM model. However, one of the most serious challenges, i.e. the Hubble constant tension, might indicate the necessity of extensions for the standard $\Lambda$CDM. In proposed extensions, dynamical dark energy or evolution of dark energy with respect to redshift have been widely discussed. Therefore, in this sense, improvement (by a factor of 2) of constraints on the dark energy equation of state is of utmost importance for shedding light on the nature of dark energy.

		%\iffalse
		\begin{table}
			\begin{center}{\scriptsize
					\caption{The $1\sigma$ uncertainties of the dark energy parameters for CPL parameterization constrained from different number of lensed GW and EM events considered.}\label{tab1}
					\renewcommand\arraystretch{1.8}
					\begin{tabular}{p{4cm}p{3cm}p{0.5cm}}
						\hline \hline 
						number of lensed GW+EM events    & $\sigma_{w_0}$  & $\sigma_{w_a}$ \\
						\hline
						0 (only SNe Ia+CMB)                         & 0.148 & 0.563 \\
						\hline
						30                          & 0.085 & 0.305 \\
						\hline
						70                          & 0.070  & 0.239\\
						\hline
						100                         & 0.063 & 0.212 \\
						\hline \hline 
					\end{tabular}}
				\end{center}
			\end{table}
			%\fi
			
\section{Conclusions and discussions}
            The strong lensing effect is an effective tool and several methods relating to this effect have been proposed for cosmological exploration. In the early days, cosmological parameters were mainly constrained by statistically comparing empirical distribution of observed image separations with the theoretical predicted one \citep{chae02,oguri12} or analyzing a large number of images of dozens of sources lensed by a single galaxy cluster \citep{paczynski1981,sereno2002,meneghetti2005,gilmore2009,jullo2010}. Later, the angular distance ratio derived from the velocity dispersion of the gravitational lens were proposed to infer the parameter information \citep{biesiada2006,grillo2008,biesiada2010,cao2012,cao2015}. Moreover, the distance ratio from strong lensing observations together with other distance measurements can also be used as an independent tool to test FLRW metric and  constrain the curvature of the universe \citep {rasanen2015}. In this distance ratio method, a relative simple profile, i.e. singular isothermal spherical or singular isothermal elliptical profile, is usually used to characterize the lens mass distribution of all measured systems. This treatment might lead to non-negligible systematic uncertainties. In addition to the above-mentioned methods, time delay measurements of strong gravitational lensing systems where each lens is individually modeled and therefore systematical bias can be significantly decreased have been gradually applied for cosmological investigations. Especially, in the most popular CPL scenario, time delay is an effective cosmological probe with different $w_0,~w_a$ degeneracy direction from those of other currently popular probes. This virtue is very helpful to precisely constrain the dark energy evolution. 
			
			The detection of gravitational waves opened a new window to observe the universe. Using GW as standard sirens for cosmology has been widely discussed in the literature \citep{schutz1986, nissanke2010, pozzo2012, cai2017, tamanini2016}. This method requires large number of GW events. When the average uncertainty of luminosity distance measurements is $10\%$, more than $10^4$ GW detections are needed to constrain the dark energy parameters with considerable precision. Thus, we combine gravitational wave and the time delay method, that is, time delay of lensed GW systems to constrain the dark energy equation of state. Because GW from DCO merging is a transient event, the time delay detection of such event is very accurate. Simultaneously, we also combine the detection of electromagnetic counterparts of GWs which enables us not only to determine the redshift of the host galaxy of the GW source, but also to improve the reconstruction of its Fermat potential difference. Therefore, these advantages are beneficial for time delay distance measurements and are further helpful for constraining the dark energy equation of state.
			
			In this paper, we quantify the complementarity of time delay of upcoming strongly lensed GW systems to currently popular probes, i.e. SNe Ia and CMB. Previous studies suggested that the third-generation detectors, i.e. ET, is expected to detect 50 to 100 lensed GW events per year, some of which will be accompanied by the observations of electromagnetic counterparts. We find that adding time delay of only 30 strong lensed GW systems to SNe Ia and CMB can improve the dark energy figure of merit by a factor 2, which is comparable to the achievement from a few $\times~10^4$ GWs with average uncertainty of luminosity distance measurements being $10\%$. In the era of precision cosmology, this improvement is of great significance for studying the nature of dark energy. Actually, there are many other applications of the lensed GW+EM events, such as testing the theory departure from the general relativity \citep{yang2018} and studying the dark matter substructure \citep{keeton2009, liao2018}.  Therefore, a large number of strongly lensed GW detections are expected in the future for exploring these interesting topics.

\section*{Acknowledgements}

We are grateful to Xuheng Ding for helpful discussions. We also would like to thank the referee for his constructive comments that have allowed us to improve the manuscript significantly. This work was supported by the National Natural Science Foundation of China under Grants Nos. 11505008, and 11633001, the Strategic Priority Research Program of the Chinese Academy of Sciences, Grant No. XDB23040100.

%%%%%%%%%%%%%%%%%%%%%%%%%%%%%%%%%%%%%%%%%%%%%%%%%%



%\bibliographystyle{mnras}

\begin{thebibliography}{99}
\expandafter\ifx\csname natexlab\endcsname\relax\def\natexlab#1{#1}\fi
				
				\bibitem[Abbott et al. (2016a)]{abbott2016a}Abbott, B. P., Abbott, R., Abbott, T. D., et al.  \ 2016, \prl, 116, 061102
				
				\bibitem[Abbott et al. (2016b)]{abbott2016b}Abbott, B. P., Abbott, R., Abbott, T. D., et al. \ 2016, \prl, 116, 241103
				
				\bibitem[Abbott et al. (2017a)]{abbott2017a}Abbott, B. P., Abbott, R., Abbott, T. D., et al. \ 2017, \prl, 121, 129901
				
				\bibitem[Abbott et al. (2017b)]{abbott2017b}Abbott, B. P., Abbott, R., Abbott, T. D., et al. \ 2017, \prl, 119, 141101
				
				\bibitem[Abbott et al. (2017c)]{abbott2017c}Abbott, B., Bloemen, S., Canizares, P., et al. \ 2017, \prl, 119, 161101
				
				\bibitem[Albrecht et al. (2006)]{albrecht2006}Albrecht, A. et al., arXiv:astro-ph/0609591
				
				\bibitem[Baker \& Trodden (2017)]{baker2017}Baker, T., \& Trodden, M. \ 2017, \prd, 95, 063512
				
				\bibitem[Bernstein et al. (2012)]{bernstein2012}Bernstein, J. P., Kessler, R., Kuhlmann, S. E., et al. \ 2012, \apj, 753, 152
				
				\bibitem[Biesiada (2006)]{biesiada2006}Biesiada, M. \ 2006, \prd, 73, 023006
				
				\bibitem[Biesiada et al. (2010)]{biesiada2010}Biesiada, M., Piorkowska, A., \& Malec, B. \ 2010, \mnras, 406, 1055
				
				\bibitem[Biesiada et al. (2014)]{biesiada2014}Biesiada, M., Ding, X., Piorkowska, A., \& Zhu, Z. \ 2014, \jcap, 10, 080
				
				\bibitem[Bonvin et al. (2017) ]{bonvin17}Bonvin, V., et al. \ 2017, \mnras, 465, 4914 
				
				\bibitem[Bruce et al. (2012)]{bruce2012}Bruce, A., Yabebal, F., Renee, H., \& Jacques, K. \ 2012, arXiv:0906.0993

               \bibitem[Cai et al. (2017)]{cai2017}Cai, R.-G \& Yang, T. \ 2017, \prd, 95, 044024
				
				\bibitem[Cao et al. (2012)]{cao2012}Cao, S., Pan, Y., Biesiada, M., Godlowski, W., \& Zhu, Z. \ 2012, \jcap, 3, 16
				
				\bibitem[Cao et al. (2015)]{cao2015}Cao, S., Biesiada, M., Gavazzi, R., Piorkowska, A., \& Zhu, Z. \ 2015, \apj, 806, 185
				
				\bibitem[Cao et al. (2014)]{cao2014}Cao, Z., Li, L., \& Wang, Y. \ 2014, \prd, 90, 062003
				
				\bibitem[Chae et al. (2002)]{chae02}Chae, K.-H., Biggs, A. D., Blandford, R. D., et al. \ 2002, \prl, 89, 151301
				
			   \bibitem[Chen et al. (2018)]{lu2018}Chen, L., Huang, Q., \& Wang, K. \ 2018, \jcap, 02, 028
				
				\bibitem[Chevalier \& Polarski (2001)]{chevalier2001}Chevallier, M., \& Polarski, D. \ 2001, IJMPD, 10, 213
				
				\bibitem[Choi et al. (2007)]{choi2007}Choi, Y., Park, C., \& Vogeley, M. S. \ 2007, \apj, 658, 884
				
				\bibitem[Collett \& Bacon (2017)]{collett2017}Collett, T. E., \& Bacon, D. \ 2017, \prl, 118, 091101
				
				\bibitem[Cristian et al. (2018)]{cristian18}Cristian, E. R., et al. \ 2018, \mnras, 477, 5657
				
				\bibitem[Ding et al. (2015)]{ding2015}Ding, X., Biesiada, M., \& Zhu, Z. \ 2015, \jcap, 12, 006
				
				\bibitem[Dominik et al. (2013)]{dominik2013}Dominik, M., Belczynski, K., Fryer, C. L., et al. \ 2013, \apj, 779, 72
				
				\bibitem[Dossett et al. (2011)]{dossett2011}Dossett, J., Moldenhauer, J., \& Ishak, M. \ 2011, \prd, 84, 023012
				
				\bibitem[Fan et al. (2017)]{fan2017}Fan, X.-L., Liao, K., Biesiada, M., et al. \ 2017, \prl, 118, 091102
				
				\bibitem[Gilmore \& Natarayan (2009)]{gilmore2009}Gilmore, J. B., \& Natarajan, P. \ 2009, \mnras, 396, 354
				
				\bibitem[Grillo et al. (2008)]{grillo2008}Grillo, C., Lombardi, M., \& Bertin, G. \ 2008, \aa, 477,397
				
				\bibitem[Jullo et al. (2010)]{jullo2010}Jullo, E., Natarajan, P., Kneib, J., Daloisio, A., Limousin, M., Richard, J., \& Schimd, C. \ 2010, Science, 329, 924
				
				\bibitem[Keeton \& Moustakas (2009)]{keeton2009}Keeton, C. R., \& Moustakas, L. A. \ 2009, \apj, 699, 2
				
				\bibitem[Li \& Li (2014)]{li2014}Li, C., \& Li, L. \ 2014, Science China, 57, 1390
				
				\bibitem[Li \& Ostriker (2002)]{li2002}Li, L., \& Ostriker, J. P. \ 2002, \apj, 566, 652
				
				\bibitem[Li et al. (2018)]{li2018}Li, Z., Gao, H., Ding, Xu., Wang, G., \& Zhang, B. \ 2018, Nature Communications, 9, 3933
				
				\bibitem[Liao et al. (2015)]{liao2015}Liao, K., Treu, T., Marshall, P., et al. \ 2015, \apj, 800, 11  
				
				\bibitem[Liao et al. (2017)]{liao2017}Liao, K., Fan, X., Ding, X., Biesiada, M., \& Zhu, Z. \ 2017, Nature Communications, 8, 1148  
				
				\bibitem[Liao et al. (2017)]{liao2017}Liao, K., Li, Z., Wang, G., \& Fan, X. \ 2017, \apj, 839, 70
				
				\bibitem[Liao et al. (2018)]{liao2018}Liao, K., Ding, X., Biesiada, M., et al. \ 2018, \apj, 867, 69
				
				\bibitem[Linder (2003)]{linder2003}Linder, E. V. \ 2003, \prd, 68, 083503
				
				\bibitem[Linder (2004)]{linder2004}Linder, E. V. \ 2004, \prd, 70, 043534
				
				\bibitem[Linder (2011)]{linder2011}Linder, E. V. \ 2011, \prd, 84, 123529
				
				\bibitem[Meneghetti et al. (2005)]{meneghetti2005}Meneghetti, M., Bartelmann, M., Dolag, K., et al. \ 2005, \aa, 442, 413
				
				\bibitem[Nakamura (1998)]{nakamura1998}Nakamura, T. \ 1998, \prl, 80, 1138

               \bibitem[Nissanke et al.(2010)]{nissanke2010}Nissanke, S., Holz, D.E., Hughes, S.A. et al. \ 2010, \apj, 725, 496
				
				\bibitem[Oguri \& Marshall (2010)]{oguri2010}Oguri, M., \& Marshall, P. J. \ 2010, \mnras, 405, 2579
				
				\bibitem[Oguri et al. (2012)]{oguri12}Oguri, M., Inada, N., Strauss, M. A., et al. \ 2012, \aj, 143, 120
				
				\bibitem[Paczy\'{n}ski \& G\'{o}rski (1981)]{paczynski1981}Paczynski, B., \& Gorski, K. \ 1981, \apjl, 248, L101
				
				\bibitem[Perlmutter et al. (1999)]{perlmutter1999}Perlmutter, S., Aldering, G., Goldhaber, G., et al. \ 1999, \apj, 517, 565
				
				\bibitem[Pi\'{o}rkowska et al. (2013)]{piorkowska2013}Piorkowska, A., Biesiada, M., \& Zhu, Z. \ 2013, \jcap, 10, 022
				
				\bibitem[Planck Collaboration (2018)]{planck2018}Planck Collaboration, arXiv:1807.06209

               \bibitem[Del Pozzo (2012)]{pozzo2012}Del Pozzo, W. \ 2012, \prd, 86, 043011
				
				\bibitem[R\"{a}s\"{a}nen \& Bolejko (2015)]{rasanen2015}R\"{a}s\"{a}nen, S. S., Bolejko, K., \& Finoguenov, A. \ 2015, \prl, 115, 10301
				
				\bibitem[Resfdal (1964)]{resfdal1964}Refsdal, S. \& Bondi, H. \ 1964, \mnras, 128, 295
				
				\bibitem[Riess et al. (1998)]{riess1998}Riess, A. G., Filippenko, A. V., Challis, P. M., et al. \ 1998, \aj, 116, 1009
				
				\bibitem[Schneider et al. (1992)]{schneider1992}Schneider, P., Ehlers, J., \& Falco, E. E. \ 1992, Berlin:
				Springer-Verlag
				
				\bibitem[Schutz (1986)]{schutz1986}Schutz, B. F. \ 1986, \nat, 323, 310
				
				\bibitem[Sendra et al. (2011)]{sendra2011}Sendra, I., Lazkoz, R., \& Benitez, N. arXiv:1105.4943 
				
				\bibitem[Sereno (2002)]{sereno2002}Sereno, M. \ 2002, \aap, 393, 757
				
				\bibitem[Sereno et al. (2011)]{sereno2011}Sereno, M., Jetzer, P., Sesana, A., \& Volonteri, M. \ 2011, \mnras, 415, 2773 
				
				\bibitem[Suyu et al. (2010)]{suyu2010}Suyu, S. H., Marshall, P., Auger, M. W., et al. \ 2010, \apj , 711, 201
				
				\bibitem[Takahashi \& Nakamura (2003)]{takahashi2003}Takahashi, R., \& Nakamura, T. \ 2003, \apj, 595, 1039

                \bibitem[Tamanini et al. (2016)]{tamanini2016}Tamanini, N., Caprini, C., Barausse, E. et al. \ 2016, \jcap, 2016, 002
                
                \bibitem[Tihhonova et al. (2018)]{tihhonova18}Tihhonova, O., et al. \ 2018, \mnras, 477, 5657
				
				\bibitem[Treu (2010)]{treu2010}Treu, T. \ 2010, \araa, 48, 87
				
				\bibitem[Yang et al. (2018)]{yang2018}Yang, T., Hu, B., Cai, R.-G., \&, Wang, B. \ 2018, arXiv:1810.00164
				
				\bibitem[Yu \& Wang (2018)]{yu2018}Yu, H., \& Wang, F.-Y. \ 2018, Eur. Phys. J. C, 78, 692 
				
				\bibitem[Wang et al. (1996)]{wang1996}Wang, Y., Stebbins, A., \& Turner, E. L. \ 1996, \prl, 77, 2875
				
				\bibitem[Wang (2008)]{wang2008}Wang, Y. \ 2008, \prd, 77, 123525
				
				\bibitem[Wei \& Wu (2017)]{wei2017}Wei, J.-J., \& Wu, X.-F. \ 2017, \mnras, 472, 2906
				
				\iffalse\bibitem[Zhao et al. (2011)]{zhao2011}Zhao, W., Den Broeck, C. V., Baskaran, D., \& Li, T. G. \ 2011, \prd, 83, 023005 \fi
\end{thebibliography}


%%%%%%%%%%%%%%%%%%%%%%%%%%%%%%%%%%%%%%%%%%%%%%%%%%

%%%%%%%%%%%%%%%%% APPENDICES %%%%%%%%%%%%%%%%%%%%%

%\appendix



%%%%%%%%%%%%%%%%%%%%%%%%%%%%%%%%%%%%%%%%%%%%%%%%%%

% Don't change these lines
\bsp	% typesetting comment
\label{lastpage}
\end{document}